\documentstyle[aps,preprint,prl]{revtex}
\begin{document}
\draft
\preprint{MIC Preprint}

\title{Coulomb Drag as a Probe of Coupled Plasmon Modes
in Parallel Quantum Wells}

\author{Karsten Flensberg and Ben Yu-Kuang Hu}

\address{Mikroelektronik Centret,
%Bygning 345{\o},
Danmarks Tekniske Universitet,\\
DK-2800 Lyngby, Denmark}

\date{June 10, 1994}
\maketitle

\begin{abstract}
We show theoretically that the Coulomb drag rate between
two parallel quasi-two-dimensional electron gases
is substantially enhanced
by the coupled acoustic and optic plasmon modes of the system at
temperatures $T \gtrsim 0.2T_F$ (where $T_F$ is the Fermi temperature)
for experimentally relevant parameters.
The acoustic mode causes a sharp upturn in the
scaled drag rate as a function of temperature at $T \approx 0.2 T_F$.
Other experimental signatures of plasmon-dominated drag are a
$d^{-3}$ dependence on the well separation $d$,
and a peak in the drag rate as a function of relative carrier densities
at matched Fermi velocities.
\end{abstract}


\narrowtext

Coulomb interactions are responsible
for a rich variety of phenomena in low-dimensional systems; {\it e.g.},
Wigner crystalization and the Fractional Quantum Hall effect.
%and possibly even the breakdown of Fermi liquid theory.
However, a direct linear-response transport measurement of the effects of
electron--electron ({\it e-e}) interactions in single isolated
semiconductor samples is usually difficult because
{\it e-e} collisions conserve crystal
momentum and hence do not degrade the current.
It has been proposed\cite{pogr77,pric83} that when
two independent electron gases with separate
electrical contacts are placed in close proximity and
current is driven through one, the inter-layer {\it e-e} interaction
creates a frictional force which drags a current through the other.
The magnitude of this drag force
is a direct measure of the inter-layer {\it e-e} interactions.
Drag responses between two- and three-dimensional
electron gases (DEG's),\cite{solo89} and between
pairs of 2DEG's in {\it e-e}\cite{gram91,gram93} and electron-hole
systems\cite{siva92} have been seen experimentally, and these have inspired
many theoretical studies.\cite{tso92,tso93,jauh93,zhen93,dragcomb}
In most experiments, no net current is allowed to flow in the layer being
dragged and a voltage arises due to charge accumulation which balances the
drag-induced carrier drift.  Measurement of this voltage, which
is directly proportional to the inter-layer drag rate, thus
provides a convenient, direct and sensitive probe of these important
inter-layer {\it e-e} interactions.

Physically, the drag arises from an inter-well transfer of momentum
caused by the inter-layer {\it e-e} interactions.
These interactions cause scattering processes in
which the electrons in the driving layer (layer 1)
lose momentum $\hbar{\bf q}$ and energy $\hbar\omega$, which are gained
by the electrons in the dragged layer (layer 2).
The drag rate $\tau_D^{-1} \equiv  e E_2/\langle p_1 \rangle$,
where $E_2$ is the electric field in layer 2, and $\langle p_1 \rangle$ is
the average drift momentum in layer 1,
has been measured in electron-filled coupled quantum
wells,\cite{gram91,gram93}
for temperatures $T \lesssim 7\,$K ($\approx 0.1\,T_{\mbox{\scriptsize F}}$,
where $T_{\mbox{\scriptsize F}}$ is the Fermi temperature).
At these temperatures, $\tau_D^{-1}$ goes essentially as $T^2$,
which agrees with the theoretical low-temperature
analyses\cite{gram91,jauh93,zhen93} for
the inter-layer momentum transfer due to Coulomb interaction.
This $T^2$ behavior is analogous to the
{\it e-e} Coulomb scattering rates in a single 2DEG.\cite{2dscat}
Deviations from the $T^2$ behavior at around $T = 3\,$K were
observed, which have been attributed to virtual phonon
exchange.\cite{tso92,macdup}
However, so far experiments have not investigated effects
at higher temperatures in this particular system, and the
low-temperature theories for $\tau_D^{-1}$, when extrapolated to
higher temperatures (beyond their range of validity), do not
show any dramatic effects.

In this paper, we show that for $T \gtrsim 0.2\,T_{\mbox{\scriptsize F}}$,
$\tau_D^{-1}$ in fact has
very interesting characteristics which are caused in large part
by contributions of the coupled-well plasmons (collective modes).
We present the following results.
(1) For two identical electron gases, the
scaled drag rate, $\tau_D^{-1}/T^2$,
which is relatively constant for low temperatures,\cite{gram91,tso92,jauh93}
has a sharp upturn around $T = 0.2 T_{\mbox{\scriptsize F}}$
and rises to a maximum
at around $T=0.5 T_{\mbox{\scriptsize F}}$.
In the peak, $\tau_D^{-1}$ is
almost {\em an order of magnitude larger} than previously
obtained results from a low-temperature approximation.
In the plasmon-dominated regime, (2)
the $\tau_D^{-1}$ dependence on the (center-to-center) well separation $d$
is approximately $d^{-3}$, as compared to $d^{-4}$ in the $T\rightarrow 0$
limit,\cite{gram91,jauh93} and
(3) at fixed $T$, as the relative density of the two electron
gases is changed, there is a maximum in $\tau_D^{-1}$
when the Fermi velocities of the two gases are the same.

How do the plasmons enhance the drag rate?
A Coulomb scattering event is a many-body process, since it
occurs in the presence of many other electrons.
The scattering matrix element is therefore given by the {\em screened}
Coulomb interaction,
$W_{12}(q,\omega) \equiv V_{12}(q)/\epsilon(q,\omega)$, where
$V_{ij}(q)$ is the bare Coulomb interaction between electrons in layer
$i$ and $j$, and $\epsilon(q,\omega)$ is the
interlayer dielectric function.
The $\epsilon(q,\omega)$ can become very small (vanishing at $T=0$)
for certain $\omega(q)$, corresponding to the presence of
plasmons of the system.
At these $\omega(q)$, $W_{12}(q,\omega)$ becomes very large,
and hence the presence of
these plasmons can greatly enhance the total {\it e-e} scattering
in the system.  In charged systems, there is generally a
high frequency (optic)
mode corresponding to all charges in the system oscillating in phase.
In systems with two distinct
components ({\it e.g.}, electron--ion or electron--hole) there can be
an acoustic branch due to the charges of the two species
oscillating out of phase.  It has been predicted that
in coupled quantum wells, such an acoustic mode should
exist.\cite{dass}
Although coupled plasmon modes have been seen by Raman scattering
in 15-layer systems,
\cite{pinc86} to the best of our knowledge,
the acoustic mode has not been explicitly
seen experimentally in a two layer system. In this paper, we point out
the novel  possibility for observing the coupled collective modes
in a transport experiment.
We show that the
acoustic mode causes the sharp upturn in $\tau_D^{-1}$, and that
both acoustic and optic modes contribute substantially to the drag
rate.  Thus, the drag rate could be used as a probe for the plasmon
modes in the coupled quantum well systems.

We now describe our calculation and results.
The drag rate between carriers in two parallel quantum wells,
to lowest order in the inter-layer interaction, is given
by\cite{siva92,tso93,jauh93,zhen93}
\begin{eqnarray}
\tau_D^{-1} &=& {\hbar^2 \over 8 \pi^2 k_B T n_1 m_2}
\int_0^\infty dq\ q^3 \int_0^\infty d\omega\ \times\nonumber\\
&&\ \ {|W_{12}(q,\omega)|^2\;\mbox{Im}[\chi_1(q,\omega)]
\;\mbox{Im}[\chi_2 (q,\omega)]
\over \sinh^2(\hbar\omega/2kT)},
\label{lifesadrag}
\end{eqnarray}
where $k_{F,i}$ and $\chi_i(q,\omega)$ are the Fermi wavevector
and susceptibility (including spin degeneracy) in layer $i$, respectively.
This integral sums up the contribution of the Coulomb scattering
between electrons in layers 1 and 2
over different momentum and energy transfers, $\hbar q$ and $\hbar\omega$.
We assume electrons only occupy the lowest energy subbands in each well,
and there is negligible tunneling,
which is consistent with experiments.\ \cite{gram91}
Within the random phase approximation (RPA),
$\epsilon(q,\omega)$ can be written\cite{jauh93,zhen93}
in terms of $\chi_i(q,\omega)$ and $V_{ij}(q,\omega)$.
For the form factors of $V_{ij}$, we assume the wells are square,
with hard-wall potentials with width $w$.

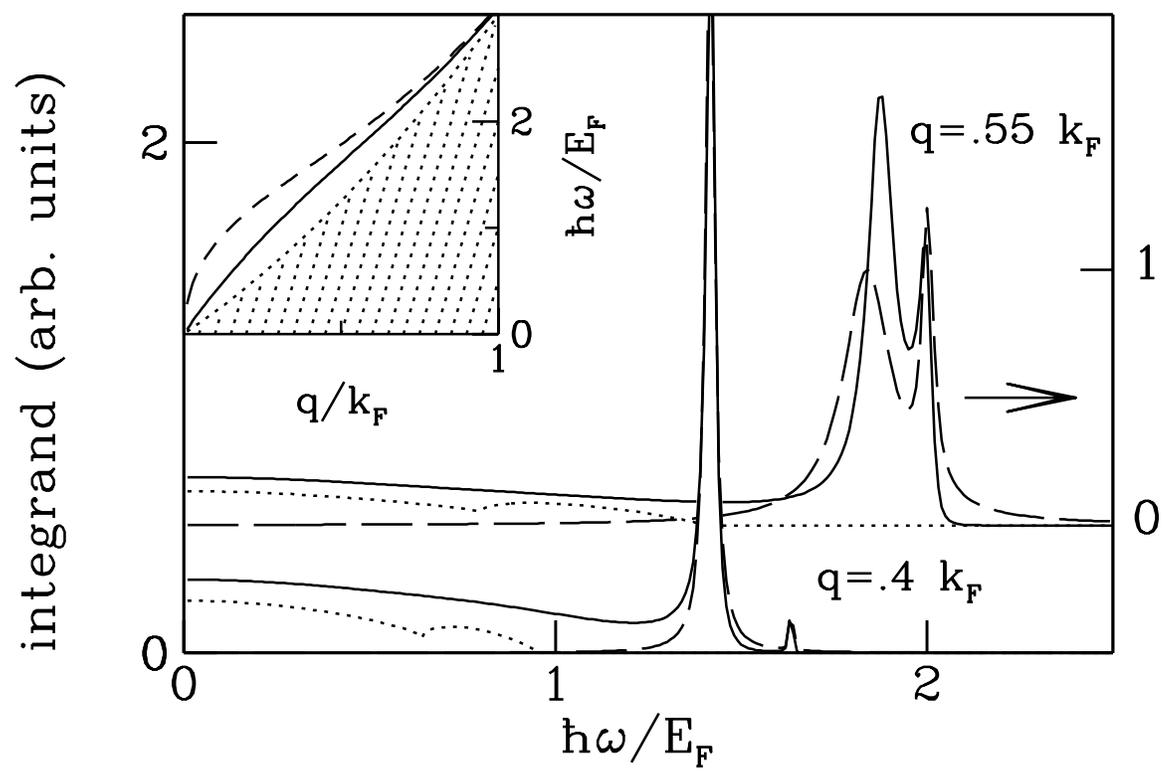
\begin{figure}
\caption{
Scattering rates [integrand of Eq.\ (\protect{\ref{lifesadrag}})] as a
function of energy transfer $\hbar\omega$, at
$q = 0.4\,k_F$ and $q = 0.55\,k_F$, for GaAs coupled quantum wells with
equal electron densities of $n = 1.5\times 10^{11}\,\mbox{cm}^{-2}$.
The well separation is $d = 3.75\, k_F^{-1} = 375\,\mbox{\AA}$, and
the width $w=0$ (except for $w$, the
parameters correspond to the experiment of Gramila
{\it et al.}\protect{\cite{gram91}}).
The solid (dotted) lines are for scattering rates
evaluated using the finite-$T$ ($T = 0$ approximation)
form of $\chi(q,\omega)$.
The dashed lines are the plasmon-pole approximation to the plasmon
peaks.  Inset: particle-hole continuum
$\mbox{Im}[\chi(q,\omega;T\!\!=\!0)]\ne 0$, (hatched area),
and acoustic and optic plasmon modes (solid and dashed lines,
respectively).
}
\label{pedagog}
\end{figure}

Previous calculations,\cite{gram91,tso93,jauh93,zhen93} which concentrated
on the low-$T$ drag rate in coupled quantum wells, have used
the $T=0$ form of $\chi_i(q,\omega)$.\cite{ster67}
(Only Sivan {\em et al.}\cite{siva92}
utilized the finite-$T$ form of $\chi$ and alluded to a plasmon
contribution in low-density systems.)
However, this $\chi_i(q,\omega; T\!\!=\!0)$ approximation
neglects the plasmon contribution because
the plasmon modes always
lie outside the $T=0$ particle-hole continuum
(as shown in the inset of Fig.\ \ref{pedagog}). In this region,
where $\mbox{Im}[\chi(q,\omega;T\!\!=\!0)]=0$
the integrand in Eq.\ (\ref{lifesadrag}) vanishes.
In contrast, for $T\ne 0$, the thermally activated
electrons above the Fermi surface make $\mbox{Im}[\chi(q,\omega)]$
finite outside the particle-hole continuum,
resulting in the plasmon enhancement of $\tau_D^{-1}$.
It is therefore essential to use the $T\ne 0$ form of
$\mbox{Im}[\chi(q,\omega)]$, and
to this end, we have developed a
numerically efficient way of evaluating the full finite-$T$ RPA form
of $\chi(q,\omega)$ arbitrary $q$ and $\omega$,\cite{flen95b}
which we use to evaluate Eq.\ (\ref{lifesadrag}).

The contributions to $\tau_D^{-1}$ can roughly be divided into
two categories: the single-particle part (where
$\mbox{Im}[\chi(q,\omega)]$ is large) and the plasmon contribution
(where $\mbox{Re}[\epsilon(q,\omega)] = 0$).
For small temperatures, $T \lesssim 0.2\, T_{\mbox{\scriptsize F}}$,
virtually the
entire contribution to $\tau_D^{-1}$ comes from the single-particle
part, but as $T$ increases above $0.2\,T_{\mbox{\scriptsize F}}$,
the plasmon contribution becomes increasingly important.
The relative contribution of these two parts is depicted in
Fig.\ \ref{pedagog}, which shows the integrand of Eq.\ (\ref{lifesadrag}),
as a function of $\omega$ for fixed $q$.
The relatively flat parts of these curves which extend from $\omega= 0$
are the single-particle contribution to the drag rate,
while the two peaks are the plasmon contributions.
The peak which is lower (higher) in frequency is the acoustic (optic) plasmon.
A quantitative estimate of the plasmon contribution can be obtained
by using a plasmon-pole approximation, which assumes that the
plasmon peaks are Lorentzian.  As shown Fig.\ \ref{pedagog},
for small $q$ and $T$, plasmon-pole approximation (dashed lines)
is a good approximation.  Even at larger $q$ and $T$, when the increased
damping makes the plasmon-pole approximation less accurate,
it still approximates the total integrated weight well and hence
is a good estimate of the total plasmon contribution to the drag rate.
Note that as temperature increases above approximately
$0.5\, T_{\mbox{\scriptsize F}}$, the plasmon peaks get extremely
broad and merge into the single-particle continuum,
blurring the distinction between the
plasmon and single-particle contributions.

Fig.\ \ref{scatrate} shows the scaled drag rate, $T^{-2}\tau_D^{-1}(T)$,
for parameters corresponding to the experiment by
Gramila {\it et al.},\cite{gram91} calculated using both
the finite-$T$ $\chi(q,\omega)$ and
the $\chi(q,\omega; T\!\!=\!0)$ approximation.\cite{zhen93}
The finite-$T$ $\chi(q,\omega)$ curve diverges sharply
at $T \approx 0.2\, T_{\mbox{\scriptsize F}}$
from the $\chi(T\!\!=\!0)$ curve.  As can be seen from
the decomposition of the drag rate into components,
this upturn is due in large measure to the {\em acoustic} plasmon,
since the acoustic plasmon is lower in energy than the
optic mode and hence is easier to excite thermally.
As the temperatures increases,
the optic plasmon also starts to contribute substantially.
As $T$ is raised further, Landau damping erodes
the oscillator strength of both modes, and their contribution diminishes,
causing the scaled drag rate to peak at approximately $T =
0.5\,T_{\mbox{\scriptsize F}}$.
At extremely large temperatures, where the system is
non-degenerate and the screening is negligible,
$\tau_D^{-1} \sim T^{-3/2}$,\cite{flen95b}
but this only occurs at $T \gg 10 T_{\mbox{\scriptsize F}}$.

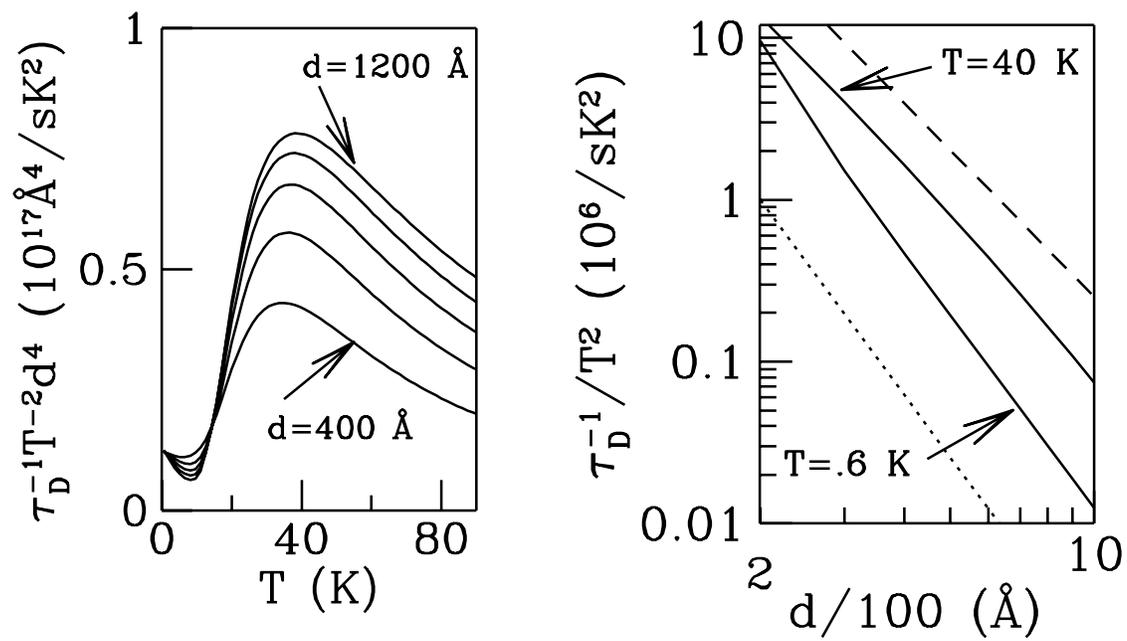
\begin{figure}
\caption{
(a) Scaled drag rates $\tau_D^{-1} d^4/T^2$,
as a function of temperature, for different well
separation $d$.  Parameters used are as in Fig.\ \protect{\ref{pedagog}},
but now with $w = 200\,\mbox{\AA}$.
The lowest curve is for $d=400\,$\AA, the highest for $d=1200\,\mbox{\AA}$
and successive curves in between differ by $200\,$\AA.
(b) Dependence of $\tau_D/T^2$ on $d$ for $T = 0.6\,\mbox{K}$
and $T = 40\,$K.
The dashed and dotted lines are reference $d^{-3}$ and
$d^{-4}$ curves, respectively.
The drag rate at $T = 40\,$K, in the plasmon-dominated regime,
has approximately an $d^{-3}$ dependence for
$d\ \protect{\lesssim}\ 700\,\mbox{\AA}$,
compared to the $d^{-4}$ dependence at $T = 0.6\,\mbox{K}$
in the single-particle dominated regime.
}
\label{ddepn}
\end{figure}

{\em Dependence of $\tau_D^{-1}$ on $d$} ---
When there is negligible plasmon enhancement
(for $T \lesssim 0.2\, T_{\mbox{\scriptsize F}}$),
$\tau_D^{-1} \propto d^{-4}$.\cite{gram91,jauh93}
This fast fall-off with $d$ is attributable to a combination of
the quasi-static ({\it i.e.,} $\omega \ll q v_F$)
screening of the system, which reduces the scattering rate
at small $q$, and the interlayer Coulomb matrix element, which
cuts off contributions for $q \gg d^{-1}$.
In the regime when the plasmon enhancement dominates, however,
we find that, $\tau_D^{-1}$ approximately has a $d^{-3}$ dependence.
This slower fall-off in $\tau_D^{-1}$ than in the low-$T$,
single-particle scattering dominated regime is to be expected,
since the quasi-static screening which contributes
to the fast $d^{-4}$ fall-off no longer applies; in fact,
the plasmon enhancement can roughly be thought of as an
anti-screening of the Coulomb interaction.
(It is difficult, however, to extract an exact analytic expression
for the $d$-dependence at plasmon-dominated temperatures.)
In Fig.\ \ref{ddepn}(a), we show $d^4 T^{-2} \tau_D^{-1}(T)$ for different
$d$.  At low $T$, the curves converge at the same point, showing the
$d^{-4}$ scaling. However, the relative size of the peak grows
larger with increasing $d$, showing a slower fall-off
than $d^{-4}$.  Fig.\ \ref{ddepn}(b) explicitly shows the
$d^{-3}$ fall-off in $\tau_D^{-1}$ in the plasmon-dominated regime.

{\em Dependence of $\tau_D^{-1}$ on relative densities} ---
The $\tau_D^{-1}$ shows a peak when the two
Fermi velocities $v_{F,i}$ are equal
(as compared to $k_{F,1} = k_{F,2}$
for phonon-mediated drag\cite{gram93}).
Fig.\ \ref{n2n1} shows peaks in $\tau_D^{-1}$
at matched densities for identical wells, and at $n_2/n_1 =
(m^*_2/m^*_1)^2$ when the effective masses $m^*_i$
in the two wells are different.
The reason for this is as follows.
Since $\tau_D^{-1}$ depends on the product
$\mbox{Im}[{\chi_1}]\,\mbox{Im}[\chi_2]$ (see Eq.\ (\ref{lifesadrag})),
the plasmon modes must be close in energy to the particle-hole
continua of {\em both} layer 1 and 2 if the plasmons are to
enhance the scattering of the thermally excited carriers.
The plasmon modes always lie above
the $T=0$ particle--hole continuum of the electron gas with the
larger Fermi velocity (otherwise, they would be heavily Landau
damped). Thus if $v_{F,1} \gg v_{F,2}$, then the phase velocity
of the plasmons is much higher than $v_{F,2}$, making it difficult
for the scattering of particles in layer 2 to be enhanced by the
plasmons.  The optimum situation for plasmon enhancement clearly
occurs when $v_{F,1} = v_{F,2}$.
As shown in Fig.\ \ref{n2n1}, the peak at $v_{F,1} = v_{F,2}$
is most pronounced at large $d$, since the relative plasmon
enhancement increases with $d$.

Before concluding, we briefly discuss the approximations and
assumptions we have used.
First, there is evidence that the RPA breaks down when the
$r_s$ parameter is greater than $5$ for large $q$.\cite{neil93}
In the experiment in Ref.\ \onlinecite{gram91}
on which the calculations in paper are mainly based,
$r_s \approx 1.4$, and most of the contribution to
the drag rate comes from $q \lesssim 0.5 k_F$, so
RPA should still be valid here.
We have also ignored virtual phonon exchange.\cite{gram93,tso92,macdup}
This virtual phonon mechanism gives a significant contribution at
around $T = 3\,\mbox{K}$, but its relative contribution to
$\tau_D^{-1}/T^2$ decreases with increasing $T$.  Thus,
since the maximum of the plasmon enhancement occurs at approximately
$30\,{\rm K}$ the virtual phonon contribution at this temperature
should be negligible.
However, since the virtual phonon exchange is weakly dependent on
$d$, the virtual phonon exchange tends to dominate for large $d$,
but we estimate that the plasmon enhancement should not be masked
for $d k_F \lesssim 10$.

In conclusion, we have calculated the drag rate for coupled quantum
wells, taking into account the effect of plasmons through proper
treatment of the dynamical screening of the coupled
electron gases.  The plasmons of the system greatly enhance the
drag rate for temperatures above $0.2\,T_{\mbox{\scriptsize F}}$.
We have elucidated the unique experimental signatures of this
plasmon enhancement in the scaled drag rate, $\tau_D^{-1}/T^2$,
which include (1) a sharp upturn at $T\approx 0.2\, T_{\mbox{\scriptsize
F}}$ and a maximum at
$T \approx 0.5\, T_{\mbox{\scriptsize F}}$ in $\tau^{-1}_D/T^2$,
(2) a $d^{-3}$ dependence in $\tau_D^{-1}$
and (3) a peak at matched Fermi velocities of the two wells.

We thank Antti-Pekka Jauho for many stimulating and informative
discussions.  KF is supported by the Carlsberg Foundation.

%\bibliographystyle{prsty}
%\bibliography{drag}

\begin{figure}
\caption{
Temperature dependence of the drag rate divided by $T^2$, for the
same parameters as in Fig. \protect{\ref{pedagog}}.
The full bold (dotted) curve corresponds to calculations using
the finite-$T$ ($T=0$ approximation) form of $\chi(q,\omega)$.
Also shown are the plasmon-pole approximation estimates for the
acoustic plasmon (ap) and optic plasmon (op)
contributions to the $\tau_D^{-1}$, and the sum of the two.
}
\label{scatrate}
\end{figure}

\begin{figure}
\caption{
Scaled drag rates as a function of the ratio of well carrier
densities, $n_2/n_1$, for
(a) $dk_F = 4$, (b) $dk_F = 8$, (c) $dk_F = 12$.
The short dashed, long dashed, dotted and solid
lines, correspond to temperatures $T = 12, 18, 30, 60\,\mbox{K}$,
respectively.
(d) Scaled drag rate at $dk_F = 8$ at $T = 30\,$K for various
ratios of effective masses in the different wells.
The solid, short dashed and long dashed lines are for
$m^*_2/m^*_1 = $ 1, 1.2 and 1.4, respectively.
The maximum plasmon enhancement
occurs when $v_{\rm F,1} = v_{\rm F,2}$;
{\it i.e.}, $n_2/n_1 = (m^*_2/m^*_1)^2$.
}
\label{n2n1}
\end{figure}

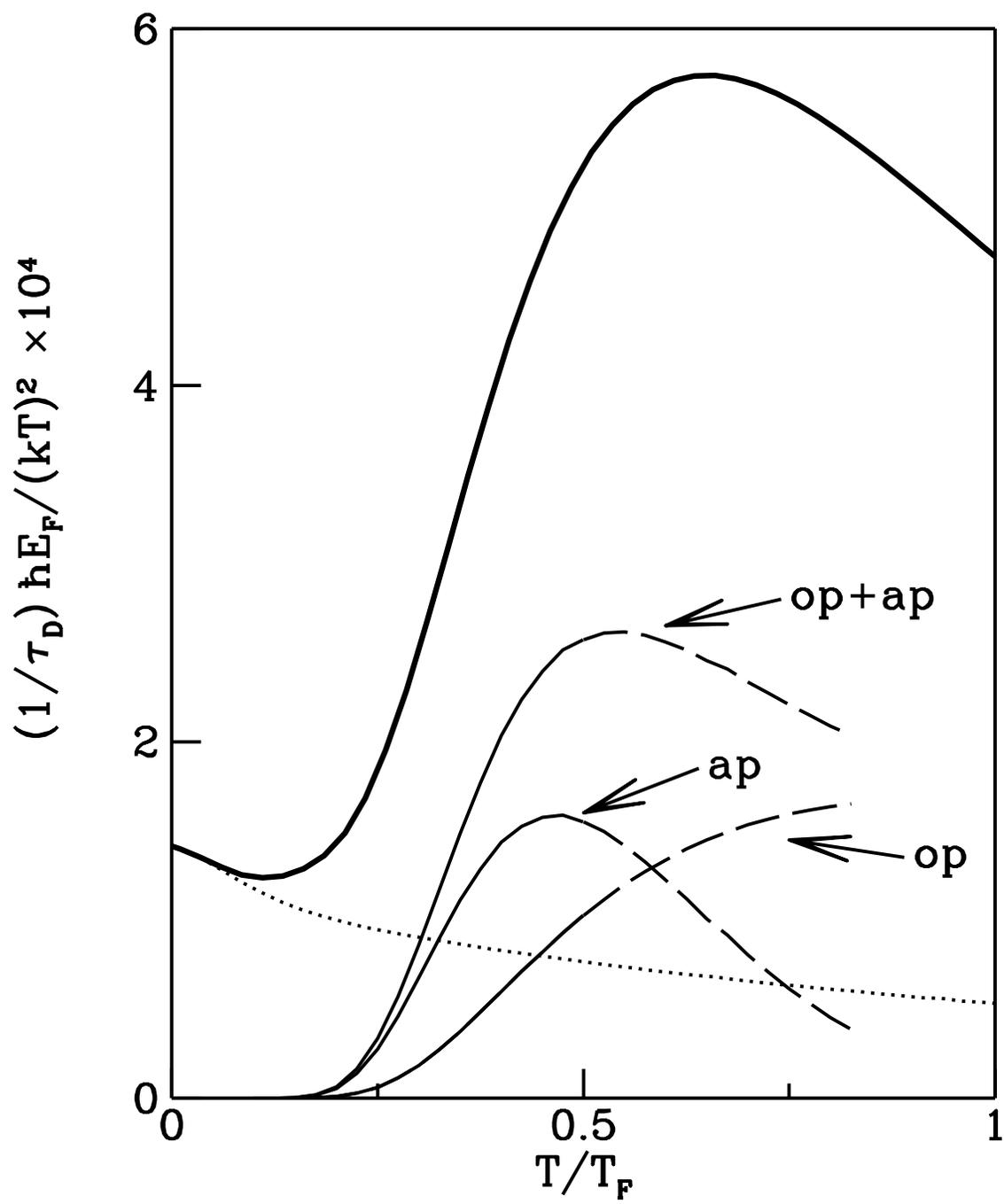
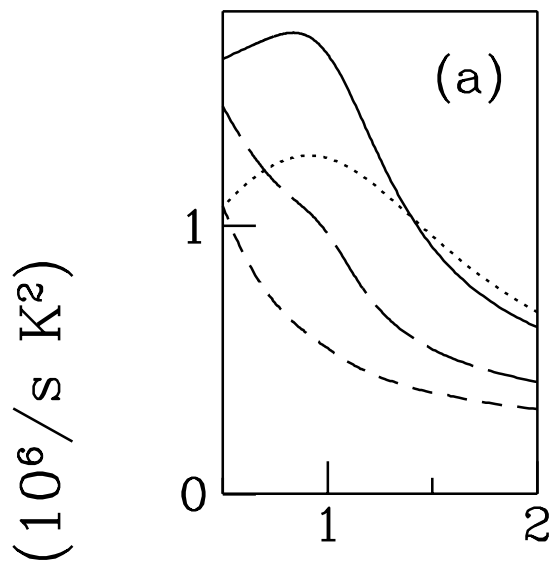
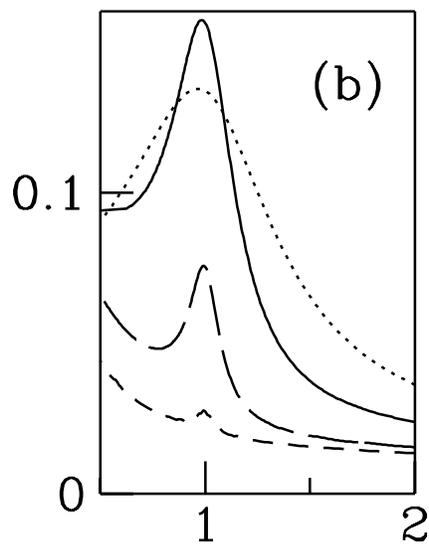
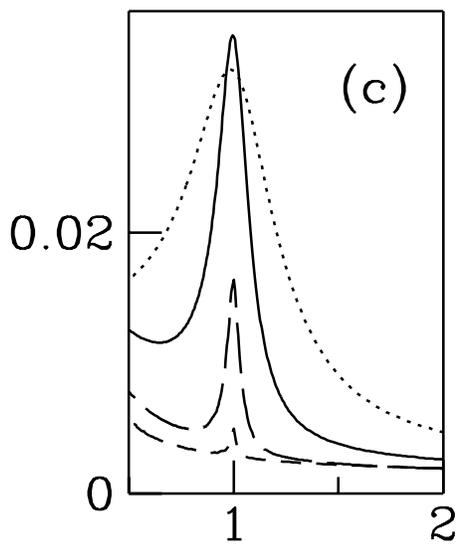
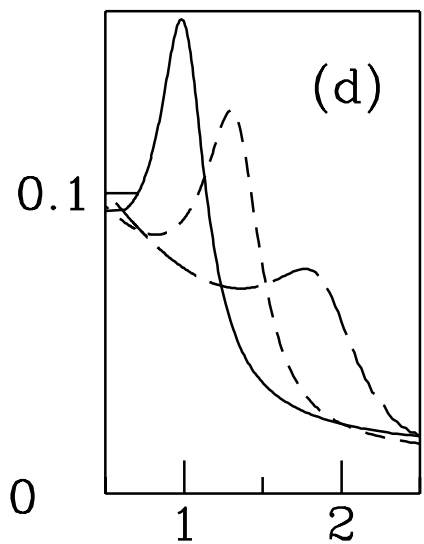
\end{document}